\documentclass[onecolumn]{aa}
\usepackage{graphicx}
\usepackage{graphics}
\usepackage{amssymb}
\def\arcmin{\hbox{$^\prime$}}
\def\deg{\hbox{$^\circ$}}
\def\ms{MS\,1054$-$0321~}
\begin{document}

\title{An X-ray review of MS\,1054-0321: hot or not? 
      \thanks{Data presented here used the XMM-{\it Newton} facility}}

   \author{I. M. Gioia
          \inst{1},
          V. Braito
          \inst{2}, 
          M. Branchesi
          \inst{1,3},
          R. Della Ceca
          \inst{2}, 
          T. Maccacaro
          \inst{2} and
          K.-V. Tran
          \inst{4}
          }

   \offprints{I. M. Gioia}

  \institute{Istituto di Radioastronomia del CNR-INAF, Via P. Gobetti 101,
              I-40129 Bologna, Italy \\
              \email{gioia@ira.cnr.it}
         \and
            INAF - Osservatorio Astronomico di Brera, Via Brera 28, 
             I-20121 Milano, Italy \\ 
              \email{braito@brera.mi.astro.it},                   
              \email{rdc@brera.mi.astro.it},                   
              \email{tommaso@brera.mi.astro.it}                   
         \and
             Dipartimento Astronomia Universit\'a di Bologna, 
             Via Ranzani 1, 
             I-40127 Bologna, Italy \\ 
              \email{mbranch@ira.cnr.it}
         \and
             Institute for Astronomy, ETH Hoenggerberg, HPF G4.2
             CH-8093 Zurich, Switzerland \\
             \email{vy@phys.ethz.ch}
}
   \date{Received ........; accepted .............}

\abstract{XMM-{\it Newton} observations are presented for the $z=0.83$
cluster of galaxies  MS\,1054$-$0321, the highest redshift 
cluster in the {\it Einstein} Extended Medium Sensitivity Survey (EMSS). 
The  temperature inferred by the XMM-{\it Newton} data, 
T$=$7.2$^{+0.7}_{-0.6}$ keV,  is much lower than the temperature 
previously reported from ASCA data, T$=$12.3$^{+3.1}_{-2.2}$ keV 
(Donahue et al. 1998), and a little lower than the Chandra temperature, 
T$=$10.4$^{+1.7}_{-1.5}$ keV, determined by Jeltema et  al. 2001.
The discrepancy between the newly derived temperature and the 
previously  derived  temperatures  is  discussed in detail. 
If one allows the column density to be a free parameter, then the best fit 
temperature becomes T$=$8.6$^{+1.2}_{-1.1}$ keV,  and the best fit 
column density becomes N$_{H}=1.33^{+0.15}_{-0.14}\times10^{20}$ 
atoms cm$^{-2}$.
The iron line  is well  detected in the  XMM-{\it Newton} spectrum  
with a value for  the abundance of Z$=$0.33$^{+0.19}_{-0.18}$ Z$_{\sun}$,
~in very good agreement with previous determinations. The derived XMM
X-ray luminosity for the overall cluster in the  2--10 keV
energy band is $L_{X}=(3.81\pm0.19) \times10^{44}$ h$^{-2}$ erg s$^{-1}$ 
while the  bolometric luminosity is 
$L_{BOL}=(8.05\pm0.40) \times10^{44}$ h$^{-2}$ erg s$^{-1}$.
The XMM-{\it Newton} data confirm the substructure in the cluster X-ray 
morphology already seen by {\emph  ROSAT} and in much more detail by Chandra.  
We find that only two of the three clumps detected in the weak lensing
mass reconstruction image are visible in X-rays, as already noted by
\cite{jel01}. The central weak lensing  clump is 
coincident  with  the  main  cluster  component and has a
temperature T$=$8.1$^{+1.3}_{-1.2}$ keV. The western weak lensing
clump coincides with the western X-ray component which is much 
cooler with a temperature T$=$5.6$^{+0.8}_{-0.6}$ keV.  
The optically  measured velocity  dispersion, obtained from  145 cluster  
redshifts, is consistent  with the velocity dispersion  expected from 
the $\sigma_{V}-T_{X}$ relationship once the XMM-{\it Newton} temperature  
is used.  \ms fits well in the  $\sigma \propto T^{1/2}$
correlation available in the literature and derived from information
collected for all clusters  at redshfits of z$\leq$1.27 known today and
with a measured X-ray temperature. The cluster temperature seems to be 
commensurate with the  predictions from its X-ray luminosity from the 
L$_{X}-T_{X}$ relation of  local  clusters. Given the newly determined 
temperature, \ms is no longer  amongst  the hottest clusters known.

}

\authorrunning{Gioia et al.}
\titlerunning{MS\,1054-0321:  hot or not?}

\maketitle


\section{Introduction}

\ms is the most luminous and distant cluster of galaxies  
in the EMSS (Gioia et al. 1990; Stocke et al. 1991).
For some time it was the only very massive, high  redshift cluster 
of galaxies (Gioia \& Luppino 1994), hence it was extensively 
studied and observed at  many  different  wavelengths. 
\cite{don98} performed an X-ray analysis using both  
ASCA and ROSAT  and determined  a high  luminosity 
(L$_{BOL}=1.1\times10^{45}h^{-2}_{100} ~erg~s^{-1}$) and  high
gas temperature (T$=$12.3$^{+3.1}_{-2.2}$ keV). Using this temperature 
Donahue et al. estimated a virial mass of $\sim7.4\times10^{14} h^{-1}_{100} 
M_{\sun}$  ~within a radius r$_{500}\footnote{within a region whose density 
is 500 times the critical density.} = 1.5 h^{-1}_{100}$ Mpc.
The X-ray emission of \ms is resolved by the ROSAT/HRI in
two, possibly  three clumps, plus an extended component indicating that 
the cluster is not relaxed and that some  caution must be adopted 
when determining  parameters from the X-ray data.
\cite{neu00} used the same ASCA temperature and an 
isothermal  $\beta$ model fit to the ROSAT/HRI data to obtain a total mass 
of $1.6\times10^{15} ~h^{-1}_{50} M_{\sun}$ within a virial radius 
r$_{V} = 1.65 ~h^{-1}_{50}$ Mpc. The significant substructure observed
in the HRI data  by the above investigators was confirmed by 
Jeltema et al. 2001 (J01) who used the superb resolution of the Chandra 
instruments. J01 give a new determination of the temperature,
T$=$10.4$^{+1.7}_{-1.5}$ keV, lower than the ASCA temperature 
and measured, consequently, a lower virial mass of 
6.2$\times10^{14} ~h^{-1}_{100} ~M_{\sun}$ within 
r$_{200}\footnote{within a region whose density
is 200 times the critical density.} = 0.76 ~h^{-1}_{100}$ Mpc. 
A similar temperature for \ms was found by \cite{joy01} using Sunyaev 
Zeldovich Effect data. They derive a temperature  T$=$10.4$^{+5}_{-2}$ keV 
and a total mass within 94\arcsec ~(corresponding to 389 kpc in the 
cosmology adopted in this paper) equal to  
M($\leq94'')=(4.6\pm0.8)\times10^{14} ~h^{-1}_{100} M_{\sun}$.  
\cite{vik02} used the Chandra instruments,  and derived an even 
lower value for the temperature, T$=$7.8$\pm$0.6 keV, in agreement 
with the determination  by \cite{toz03}  who analyzed the same 
Chandra data and obtained T$=$8.0$\pm$0.5 keV.
In the optical band an extensive study by  \cite{tran99} and \cite{vdok00}, 
based on 24 and 81 spectroscopic cluster member redshifts, respectively,
finds agreement between the observed velocity dispersion 
($\sigma_V=1170\pm$150 km s$^{-1}$; Tran et al. 1999) and the high  X-ray 
measured  ASCA temperature (T$=$12.3$^{+3.1}_{-2.2}$ keV; Donahue et al. 
1998),  albeit within the large errors.
From Hubble Space Telescope (HST) imaging, supplemented by ground
telescope data, van Dokkum et al. (2000) detected  a large overdensity
of red cluster galaxies with irregular and elongated distribution.
A very high fraction (17 per cent) of the cluster galaxies are classified 
as ``mergers/peculiar'' on the basis of double nuclei, tidal tails and 
distorted morphologies (see van Dokkum et al. 1999; van Dokkum et al. 
2000). 

Hoekstra, Franx \& Kuijken (2000)  performed a weak lensing study using a 
two-color  mosaic of deep WFPC2 images and determined  a mass of
(1.2$\pm$0.2$)\times10^{15}$ $h^{-1}_{50} M_{\sun}$ within an aperture radius
1~$h^{-1}_{50}$ Mpc, in good agreement with the previous estimators 
(Luppino \& Kaiser 1997; Clowe et al. 2000). Assuming an isothermal mass 
distribution \cite{hoek00}  found a corresponding velocity
dispersion of 1311$^{+83}_{-89}$ km s$^{-1}$, again consistent
with the X-ray and optical estimates. The weak lensing signal clearly 
shows three distinct clumps in the mass reconstruction distribution in 
agreement with the light distribution, that all appear to have similar masses. 

An extremely deep 5-GHz survey of the cluster region  down to 32 $\mu$Jy 
(6$\sigma$; Best et al. 2000) detects a high number of radio sources 
(34 vs 25 expected from blank-field radio source counts). 
Surprisingly, no radio emission  is detected towards the merger galaxies 
found in the HST images by \cite{vdok00} consistent with the hypothesis 
that low-luminosity radio sources may be triggered by initial weak 
interactions rather than direct mergers (Best et al. 2000).
\cite{joh03} detected an excess of point sources  in the
archival deep Chandra image of the field of MS\,1054$-$0321. Combined 
with identifications of cluster AGN  from the radio work of \cite{best02}, 
\ms seems to have a significantly enhanced AGN acitivity with respect 
to local galaxy clusters.

From all these data a consistent picture emerges of \ms as a young, massive, 
highly luminous cluster with significant substructure both in the  X-ray and
optical data, confirmed by the weak lensing mass reconstruction distribution.
\ms was also observed by XMM-{\it Newton}.  The new data analysis and 
interpretation  are the subject of this paper. A new redshift determination
from additional optical spectroscopic data is used in combination
with the XMM-{\it Newton} data to  provide a more accurate picture 
of  the cluster. 
Comparisons are  also made with previous analyses. Throughout the paper we 
use  a Hubbble constant of $H_{0} = 100 h$ km s$^{-1}$ Mpc$^{-1}$,  q$_{0}=0.5$
(for comparison with previous work on this cluster) and a cosmological 
constant of $\Lambda=0$, except where noted. 
One arcmin corresponds to 248.6 h$^{-1}$ kpc at the cluster redshift.

\begin{figure}
   \centering
      \caption{X-ray image of the MOS2 detector showing the extended emission
of MS\,1054$-$0321. This image is presented to highlight the four different 
regions analyzed in X-rays. Figure 1 can be seen in the pdf file of the paper
at http://www.ira.cnr.it/~gioia/PUB/publications.html}
         \label{Fig1}
   \end{figure}
%

   \begin{figure}
   \centering
   \includegraphics[width=10cm, angle=-90]{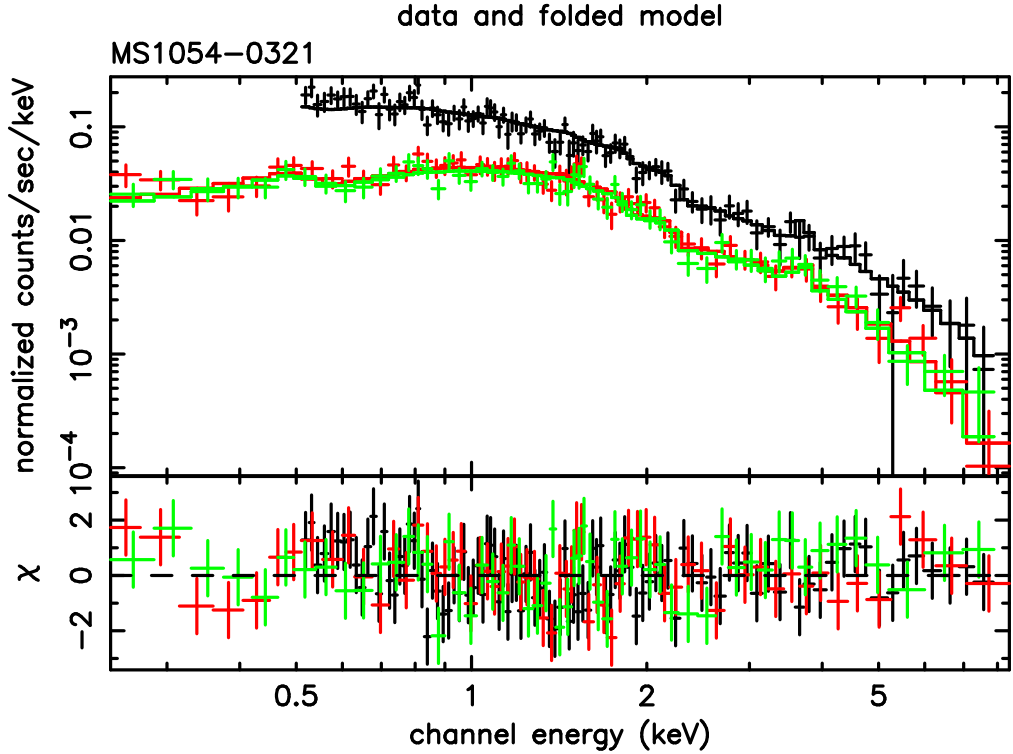}
      \caption{Binned X-ray spectrum and residuals for MS\,1054$-$0321. 
         The spectrum was binned to have a minimum of 40 (50) counts  
         per bin for each MOS (pn). The solid line represents the 
         the best fit MEKAL model.}
         \label{Fig2}
   \end{figure}

   \begin{figure}
   \centering
   \includegraphics[width=8cm, angle=-90]{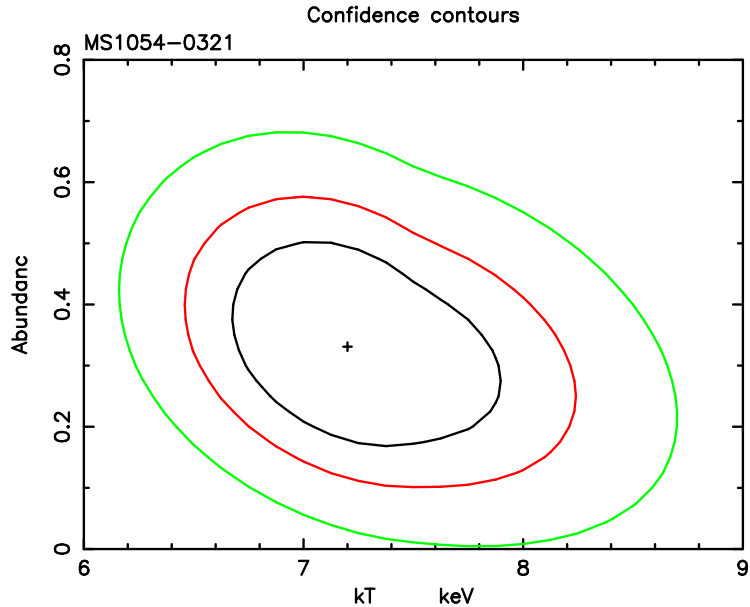}
      \caption{68.3\%, 90\%, and 99\% confidence $\chi^{2}$ contours 
       ($\Delta\chi=$2.30, 4.61, and 9.21) for the cluster iron abundance 
       and temperature (point source to the south excluded).}
         \label{Fig3}
   \end{figure}

\section{XMM-{\em Newton} Data}

\subsection{Observations and Data Analysis}

XMM-{\it Newton} observed \ms ~as part of 
the GTO program, on June 21, 2001 for about 40 Ks (Seq \# 0094800101).
The European Photon Imaging Camera (EPIC), which incorporates one pn and 
two MOS CCD arrays (Sr\"uder 2001; Turner et al. 2001) was
operating  in full-frame mode with the thin filter 
applied. Unfortunately the observations suffer from time periods of  
very high background.
Event files produced from the standard pipeline  processing have been
examined and filtered (using the version 5.4 of the Science Analysis
Software, SAS, and  the latest calibration files released by the EPIC
team) to remove the high background time intervals. Only events corresponding
to pattern 0-12 for MOS and pattern 0-4 for pn have been used\footnote{see the
XMM-{\it Newton} Users' Handbook \\ 
{\it http://xmm.vilspa.esa.es/external/xmm\_user\_support/documentation/uhb/XMM\_UHB.htm}}.
The net exposure times, after data cleaning, are  $\sim$ 28.4 Ks and 
$\sim$ 29 Ks for MOS 1 and MOS 2, respectively,
and $\sim$ 21.9 Ks for pn.  Background counts have been accumulated using 
nearby source-free circular regions.  This observation suffers from high mean 
background level thus the use of a local background estimate is more 
appropriate than the blank-sky background  which would lead to an 
underestimate of the background counts.
Response matrices (that include the correction for the 
effective area) have been generated using the SAS tasks {\it arfgen} and 
{\it rmfgen}. The X-ray fluxes reported below are computed using the
MOS2 detector and calibration(s). The normalization of the MOS1 is 
about 2\% higher than for MOS2 whereas the normalization of the pn is 7.5\% 
higher.

\subsection{Overall Cluster Properties}

Fig. \ref{Fig1} shows the XMM-{\it Newton} MOS2 image of  MS\,1054$-$0321.  
The X-ray cluster is elongated  in the east-west direction and follows  
the distribution of optical galaxies well visible in the I-band image 
of \ms shown in Fig. \ref{Fig7}.
The point source to the south is  also marked. The X-ray image 
has two distinct components (Component~1 and Component~2)  plus a 
fainter emission region (indicated as Component~3 in the figure) 
between the  cluster main  emission and the point source  emission.

In order to measure the emission-weighted  cluster temperature, net counts
were extracted from a circle of radius 1.5\arcmin ~centered at
$\alpha_{2000}= 10^{h}~56^{m}~58.7^{s},  ~\delta_{2000} = -03\deg~37\arcmin
~45\arcsec$ in the MOS1 and MOS2 detectors. The 1.5\arcmin ~radius was chosen
to have the same parameters previously used by J01 in order to compare
our results with theirs. 
There are 2392$\pm$54 net counts, in the 0.2--8 keV energy range, in  
MOS2 (2198$\pm$52 in MOS1) and 4486$\pm$75 net counts in the pn  detector 
(the pn counts are taken in the 0.5--8 keV band to avoid calibration
uncertainties of the pn in the softer energy band).
The data were fitted with a  MEKAL model (Mewe, Gronenschild \&  van den 
Oord, 1985) modified by Galactic absorption. The absorption and 
redshift were kept constant while the abundance and temperature were 
left free to vary.  The Galactic absorption was 
fixed at the Galactic value of $3.6\times10^{20}$ atoms cm$^{-2}$
(Dickey \& Lockman 1990), and the redshift was fixed at  z$=$0.83 
(Tran et al. 1999; van Dokkum et al. 2000).  The point source to the south 
(indicated as Point Source in Fig.~\ref{Fig1}) at  
$\alpha_{2000}=10^{h}~56^{m}~58.8^{s}$,
~$\delta_{2000}=-03\deg~38\arcmin~52\arcsec$, corresponds to the
point source previously seen as an extension to the south in the  ROSAT 
HRI data, and as a point source in the Chandra image published by J01.
This source was excluded from the analysis  of the cluster by masking
a circular region with radius 15\arcsec, and will be discussed later.
The best fit with a thermal MEKAL model gives a temperature 
T$=$7.2$^{+0.7}_{-0.6}$  keV and an  abundance 
Z$=$0.33$^{+0.19}_{-0.18}$ Z$_{\sun}$, relative to the abundances  
of \cite{gs98}\footnote{As suggested by  \cite{gm02}, we have adopted 
the Grevesse \& Sauval values since the community has converged towards 
a ``standard solar composition'' given the discrepancies
between meteoritic and photospheric solar abundances, particularly for
iron, as quoted in a review by \cite{and89}. We note that using the 
ratio between the elements fixed to the solar values of Grevesse \& Sauval 
results in abundance values higher by a factor 1/0.676 than using
the solar values of Anders \& Grevesse (1989). This is due to the fact 
that Grevesse \& Sauval use a 0.676 times lower iron abundance.}.
The best fit X-ray spectral parameters are listed in Table 1.
Throughout this paper quoted uncertainties are 90\% confidence levels for 
one interesting parameter. 

The binned X-ray spectrum and best fit folded model are shown in 
Fig.~\ref{Fig2} where  bins include at least 40 counts each (50
counts each for the pn).
The fit has an acceptable reduced $\chi^{2}$ of  0.986 for 219
degrees of freedom. The  $\chi^{2}$ contours for iron abundance 
versus  cluster temperature are shown in Fig.~\ref{Fig3}. The 
absorbed flux in the 2--10 keV energy band, excluding the point-like 
source to the south,  is F$_{X}=~$(3.40$\pm$0.17)$\times10^{-13}$ erg 
cm$^{-2}$ s$^{-1}$. The unabsorbed K-corrected luminosity in the same 
energy band is  $L_{X}=(3.81\pm0.19) \times10^{44}$ h$^{-2}$ erg s$^{-1}$ 
while the  bolometric luminosity is $L_{BOL}=(8.05\pm0.40) \times10^{44}$ 
h$^{-2}$ erg s$^{-1}$. Given the detection of the iron line, we can thaw 
the redshift parameter in XSPEC and fit for it. The best fit for
the redshift is z$=$0.847$^{+0.057}_{-0.040}$, in very good agreement  
with the optical spectroscopic determination.

We have then fitted the data allowing  the N$_{H}$  to be a free 
parameter and using a one-temperature MEKAL model. The resulting
value  for the column density is  
N$_{H}=1.33^{+0.15}_{-0.14}\times10^{20}$ atoms cm$^{-2}$,
lower than the Galactic value, and the resulting temperature
is  T$=$8.6$^{+1.2}_{-1.1}$ keV, higher than when freezing the 
N$_{H}$. The abundance becomes 
Z$=$0.22$^{+0.15}_{-0.14}$ Z$_{\sun}$, and the  reduced 
$\chi^{2}$  is 0.923 for 218 degreees of freedom.
This result goes in the same direction as the result by 
Tozzi et al. 2003 for \ms (see their Fig. 12) where they also 
obtain  an higher  temperature ($\sim$ 10 keV) and a lower 
column density value when they allow the N$_{H}$ to vary.

We have also explored the effect of fitting a two-temperature MEKAL model 
to the data  to check if there is a soft emission coming from a cooler 
temperature  plasma, due to the presence of groups and
poor clusters emission  falling into the larger cluster. 
We obtain a value for the column density lower
than the Galactic value: N$_{H}$ goes from 1.3$\times10^{20}$ to 
2.6$\times10^{20}$  atoms cm$^{-2}$, while the two temperatures
assume values of $\sim$0.2 keV and $\sim$9 keV but with a great
degeneracy which makes impossible to assign any error to the fit.
From a purely statistical point of view, the addition of the softer thermal 
component is not supported by the F-test (Bevington \& Robinson, 1992). 

  \begin{figure}
   \centering
\caption{X-ray spectrum and residuals for the eastern X-ray component
         (Component~1) within a 0.47\arcmin ~radius circle. The spectrum was
         binned to have a minimum of 20 counts  per bin. The solid line 
         represents the best fit MEKAL model. The insert shows the 68.3\%, 
         90\%, and  99\% confidence $\chi^{2}$ contours (corresponding to 
         $\Delta\chi^{2}=$2.30, 4.61, and 9.21) for the
         iron abundance and temperature.  Figure 4 can be seen in 
         the pdf file of the paper at 
         http://www.ira.cnr.it/~gioia/PUB/publications.html}
 \label{Fig4}
   \end{figure} 

 \begin{figure}
   \centering
\caption{X-ray spectrum and residuals for the western X-ray component 
         (Component~2) within a 0.47\arcmin ~radius circle. The spectrum was
         binned to have a minimum of 20 counts per bin. The insert shows the 
         68.3\%, 90\%, and 99\% confidence $\chi^{2}$ contours (corresponding 
         to  $\Delta\chi^{2}=$2.30,  4.61, and 9.21) for the iron abundance 
         and temperature.  Figure 5 can be seen in the pdf file of the paper
         at http://www.ira.cnr.it/~gioia/PUB/publications.html}
 \label{Fig5}
   \end{figure}

 \begin{figure}
   \centering
 \includegraphics[width=8cm, angle=-90]{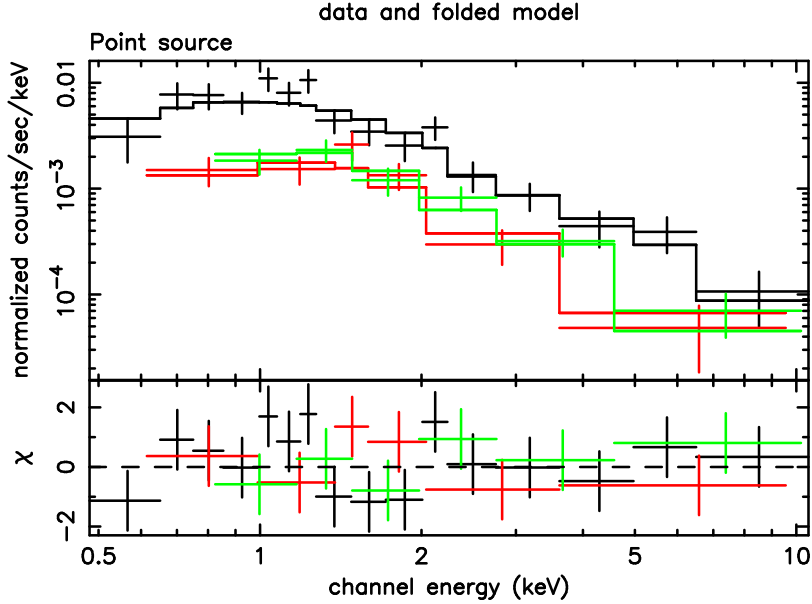}
\caption{X-ray spectrum and residuals for the point source to the south of the
          cluster  within a 0.33\arcmin ~radius circle. The spectrum was 
          binned to have a minimum of 20 counts  per bin. The solid line 
          represents the best fit power law model.}
 \label{Fig6}
   \end{figure}

\subsection{Cluster Substructure}

Jeltema et al. (2001) show in Fig. 9 of their paper that the 
central  and  western clumps of the weak lensing study correspond to 
the eastern and western Chandra X-ray clumps, respectively. Several 
galaxies  lie at the position of the eastern clump,  including the cD, 
while  the western  clump appears  to have galaxies 
which  are lining the southern edge of  the clump. 
We have enough statistics in the XMM-{\it Newton} data to examine the two X-ray peaks 
(Component~1 and Component~2) separately  by putting an extraction circle 
of radius   0.47\arcmin ~centered  on each peak. The  0.47\arcmin ~radius 
was chosen to avoid overlapping the extraction circles placed 
on the two separate clumps and thus any possible contamination 
between them (see Fig. \ref{Fig1}). For the eastern component
the center was fixed at $\alpha_{2000}= 10^{h}~57^{m}~00.3^{s},
~\delta_{2000} =-03\deg~37\arcmin~28.8\arcsec$, while the western peak 
is centered at  $\alpha_{2000}= 10^{h}~56^{m}~56.6^{s},
~\delta_{2000} =-03\deg~37\arcmin~36.3\arcsec$.
There are between 580 and 660 net counts in each clump according to
which  MOS is considered (see Table 1), and approximately over twice 
this number of 
counts in the pn. In total there are 2387$\pm$49 net counts in the
eastern clump and  2245$\pm$50 net counts in the western clump.
The spectrum of each clump was fitted with a MEKAL model with the 
N$_{H}$ fixed at the Galactic value of $3.6\times10^{20}$ atoms cm$^{-2}$,
the redshift fixed at 0.83, and the iron abundance and temperature were  
left free to vary.  Given the reduced statistics with respect to the cluster 
as  a whole, the data were binned to include a minimum of 20 counts per bin. 
The eastern component has a best-fit temperature of T$=$8.1$^{+1.3}_{-1.2}$ 
keV with a reduced $\chi^{2}$ of 1.4 for 115 degrees of freedom. 
This temperature is consistent with the cluster temperature, while  
the western component is somewhat cooler with a best-fit temperature of 
T$=$5.6$^{+0.8}_{-0.6}$ keV  and a reduced  $\chi^{2}$ of 1.26 for 
110 degrees of freedom. The spectra and best-fit models are shown in 
Fig.~\ref{Fig4} and Fig.~\ref{Fig5}. The western clump was found to be 
cooler also in the Chandra analysis by  J01 who quote a temperature 
T$=$10.5$^{+3.4}_{-2.1}$ keV for the eastern clump, and
T$=$6.7$^{+1.7}_{-1.2}$ keV for the western clump.
We find the iron abundances of the eastern and western components to be
Z$=$0.12$^{+0.35}_{-0.12}$ Z$_{\sun}$ ~and Z$=$0.51$^{+0.36}_{-0.32}$ Z$_{\sun}$, 
respectively. Given the large uncertainties these abundances are found 
to be consistent between  each other and with the cluster abundance. 
Similar results were found by J01 even if their abundances are relative 
to    \cite{and89}. The absorbed fluxes in 2--10 keV for the two clumps
are F$_{X, east}=$(9.69$\pm$0.67)$\times10^{-14}$ erg cm$^{-2}$ s$^{-1}$ and
F$_{X, west}=$(6.94$\pm$0.59)$\times10^{-14}$ erg cm$^{-2}$ s$^{-1}$, 
respectively, while the unabsorbed K-corrected luminosities in the same 
energy bands are
L$_{X, east}=$(1.05$\pm$0.07)$\times10^{44}$ erg s$^{-1}$ and
L$_{X, west}=$(8.69$\pm$0.73)$\times10^{43}$ erg s$^{-1}$. The respective
bolometric luminosities are L$_{BOL,east}=$(2.23$\pm$0.15)$\times10^{44}$ erg 
s$^{-1}$ and L$_{BOL, west}=$(1.82$\pm$0.15)$\times10^{44}$ erg s$^{-1}$.
We attempted  to examine also the emission region marked as Component~3 in 
Fig. \ref{Fig1} by putting an  extraction circle of radius 
0.47\arcmin ~centered as indicated in the figure. A higher temperature 
than the cluster as a whole is found of T$=$8.2$^{+2.45}_{-1.71}$ keV, and an 
abundance of Z$=$0.23$^{+0.63}_{-0.23}$ Z$_{\sun}$. The large errors are due 
to the  reduced statistics. There are only $\sim$350 net counts per 
detector. Given the  small number of photons, this emission region 
will not be considered  any further.

As previously noted by \cite{clo00} and by J01 the eastern weak lensing 
clump is X-ray  dark. It is  unclear why this mass density peak is not 
seen in X-rays. As J01 suggested,  \ms ~could be in a pre-merger state given 
the clear separations of the  two X-ray peaks. In this scenario the eastern 
weak lensing peak  could still be forming and not be yet virialized, and 
it is thus underluminous in X-rays.

\subsection{Point Source to the South}

In order to examine the point source to the south, a circle of 0.33\arcmin 
~was centered at $\alpha_{2000}=10^{h}~56^{m}~58.8^{s}$,
~$\delta_{2000}=-03\deg~38\arcmin~52\arcsec$. There are only 
a total of 454$\pm$24 net counts in the source. Fitting to an absorbed 
power  law gives a N$_{H}=(1.7\pm0.7)\times10^{21}$ atoms cm$^{-2}$ with 
a photon index $\Gamma$=1.9$\pm$0.2, and a reduced  
$\chi^{2}$ of 0.97  for 23  degrees of freedom. Both the column density 
(higher than the Galactic value) and the power law values are in very good
agreement with the Chandra analysis by J01, who find a power
law with a  photon index $\Gamma$=1.7$\pm$0.3, and 
a N$_{H}=2.2^{+1.8}_{-1.4}\times10^{21}$ atoms cm$^{-2}$ . 
The observed flux in the 2--10 keV energy band for this source is  
F$_{X}=~$(3.0$\pm$1.0)$\times10^{-14}$ erg cm$^{-2}$ 
s$^{-1}$. An X-ray spectrum is given in Fig.~\ref{Fig6}. 
The X-ray parameters are reported in Table 3. The  optical image shows
a quasi stellar object for which no optical spectroscopy was performed.
We do not have an identification for this source. However its point-like 
appearance in the optical image  plus the X-ray spectrum which has a photon 
index typical of Seyfert 1 active  galactic nuclei suggest that the source 
could be an AGN. 

%

\section{Optical Data} \label{optical}

Deep optical CCD images of \ms from the ground and from space have been 
acquired by several investigators in order to perform weak 
lensing studies and  morphological analyses of the cluster 
galaxy population (Clowe et al. 2000; Hoekstra, Franx \&  Kuijken 
2000; van Dokkum et al. 2000). \cite{vdok99} discover a high fraction of 
galaxy mergers in \ms which argues  against formation of elliptical  galaxies 
in a single  ``monolithic'' collapse at high redshift but  is in qualitative 
agreement with predictions of hierarchical models for structure formation.
An impressive $4'\times6'$ HST mosaic obtained with deep WFPC2 
images in two colors is published in \cite{vdok00}. The mosaic shows 
the filamentary appearence of MS\,1054$-$0321: the cluster is easily 
identified as the horizontal and broad  swath of red galaxies in the center 
of the frame. The cluster has also been  observed several times at the 
Keck telescope  to obtain spectroscopic redshifts for the member galaxies
(Donahue et al. 1998; Tran et al. 1999; van Dokkum et al. 2000). 

A larger number of spectroscopic redshifts mainly taken at the Keck telescope
of both field galaxies and cluster members have been  gathered by one of 
us (K-VT) and her collaborators.
From a magnitude selected sample Tran et al. (in preparation) have
spectroscopically confirmed cluster membership for 145 galaxies in a
7$\arcmin\times7\arcmin$ region centered on the cluster.  From this 
large sample, they  determine the cluster mean redshift and velocity
dispersion  to be z$=$0.8308$\pm$0.0006 and $\sigma_V=1153\pm$80 km 
s$^{-1}$, with smaller uncertainties with respect to previous 
determinations by these same  authors. \ms may well be the only z$=$0.8  
cluster  today with such a huge number of redshift determinations. 
The substructure evident in both X-ray and weak-lensing maps is also
observed in the optical.  The most significant optically identified 
subcluster lies 45\arcsec ~east of the BCG (Bright Cluster Galaxy) and, 
in comparison to the bulk of cluster members, is at lower  redshift 
(z$=$0.8281$\pm$0.0016) and has a smaller velocity dispersion  
($\sigma_V=1061\pm168$ km s$^{-1}$). This clump is the only significant 
clump identified using the Dressler-Shectman test (Dressler \& Shectman 1988).
A direct comparison among the clumps in X-ray, weak lensing 
and optical data is  not straightforward since the peaks do not coincide
in the three wavebands.  We have placed optical circles corresponding to 
the eastern and western X-ray clumps  for comparison   purposes. 
Redshifts and velocity dispersions have been  recomputed for the galaxies
in the two X-ray clumps. These data are presented in Table 4. 
The discrepancy in their mean redshifts and their large dispersions 
illustrates how unrelaxed \ms is and how much the distribution of the 
X-ray gas deviates from the optical distribution of galaxies.
 
\section{Results and Discussion}

The main result from the new XMM-{\it Newton} data is the lower measured 
value for the temperature of \ms with respect to previous determinations.
The XMM-{\it Newton}  temperature, T$=$7.2$^{+0.7}_{-0.6}$ keV, is
definitely lower than the ASCA temperature T$=$12.3$^{+3.1}_{-2.2}$ keV  
(Donahue et al. 1998), and also somewhat lower, even taking into
account the uncertainties, than the  Chandra  temperature, 
T$=$10.4$^{+1.7}_{-1.5}$ keV, reported by J01. 
It is however  fully consistent with the Chandra temperature,
T$=$7.8$\pm$0.6 keV, published by \cite{vik02} (see their Table 2),
and  with the Chandra temperature, T$=$8.0$\pm$0.5 keV, published by 
\cite{toz03} (see their Table 3). The iron abundance inferred from the
XMM-{\it Newton} observations is found to be in very good agreement 
with previous  determinations by  both J01 (Z$=$0.38$\pm$0.15 Z$_{\sun}$) 
and by \cite{toz03}  (Z$=$0.35$\pm$0.07 Z$_{\sun}$) after their abundances 
are rescaled to the \cite{gs98}  values while Vikhlinin et al. (2002)
fix the abundance to  Z=0.3 Z$_{\sun}$.
\ms is in many aspects very similar  to other distant X-ray 
selected   clusters like RXJ\,1717+67 at  $z=0.81$  (Gioia et al. 1999), or  
RXJ\,0152.7$-$1357 at $z=0.83$  (Della Ceca et al. 2000; Ebeling et al. 
2000) which show  elongated or double morphologies  in optical, weak 
lensing and X-rays and which have moderate temperature values.

While it is possible that ASCA included other sources in its larger field 
of view, the lower X-ray  temperature found for this cluster by this 
analysis, seems to be more  accurate given the agreement of three out of 
four measurements (this paper; Vikhlinin et al., 2002; Tozzi et al., 2003).  
In order to verify the correctness of our XMM-{\it Newton} results we have 
analyzed  the Chandra archival data of \ms ~and used all the information 
in the J01 paper to duplicate exactly their data reduction. There is 
a number of corrections to be made to the data. One of them, the
ACISABS correction\footnote{http://www.astro.psu.edu/users/chartas/xcontdir/xcont.html}, 
was not available at the time of the J01 analysis and was implemented 
later in the suggested steps for Chandra data reduction. This correction 
takes into account the quantum efficiency degradation at low energies, below 
1 keV. The application of the ACISABS correction shifts the temperature
towards lower values. There are several flares in the Chandra data, mostly
visible only in the ACIS S3 back-illuminated chip, which contribute to 
raise the temperature. In the following analysis we use: a 1.5\arcmin 
~extraction radius at the same cluster center as J01, 
the 0.8--7 keV energy range  for the  spectral fit,  subtraction 
of  point sources including the one to the south at the 
same position as J01, local background estimate, use of the Anders \& 
Grevesse abundance values, counts binned in intervals so as to 
have at least 20 counts per bin, fit with a Raymond-Smith (Raymond \& 
Smith 1977) thermal plasma  model, N$_{H}$ fixed to  the 
Galactic value, temperature and abundance left free to vary. The 
two different parameters used are a lower exposure time of 67,478s
(to cut out the time intervals affected by the flares) and application 
of the  ACISABS correction. We obtain a value for the Chandra temperature
T$=$7.4$^{+1.4}_{-0.9}$ keV,  and an abundance Z$=$0.20$\pm$0.14 Z$_{\sun}$, 
fully consistent with our XMM-{\it Newton} spectral result. If instead 
we release  the two corrections (ACISABS, lower exposure time) then we 
obtain the same values as  published by J01,  namely  
T$=$10.1$^{+1.7}_{-1.5}$ keV and  Z$=$0.22$\pm$0.16 Z$_{\sun}$. 
Adopting all the corrections and the same parameters as in \cite{vik02}
we obtain exactly their same value  for the temperature 
T$=$7.8$\pm$1 keV. Thus we are confident that our XMM-{\it Newton} 
data analysis is correct since the two different datasets, 
XMM-{\it Newton} and Chandra, yield exactly the same result.

We check now how \ms fits in the  L$_{X}-T_{X}$ relation of galaxy clusters.
Understanding the evolution of the  L$_{X}-T_{X}$ relation is important not 
only to understand the  physics behind the formation of galaxy clusters, but
also because it  provides a link between observations of clusters and
derivation  of  cosmological parameters. The L$_{X}-T_{X}$ relation has been 
well studied at low redshift. There is no consensus yet on the evolution 
of the L$_{X}-T_{X}$ relation with redshift (see among others: Borgani et 
al. 2001; Holden et al. 2002; Novicki et al. 2002; Vikhlinin et al. 2002)
probably due to the lack of large  samples of  galaxy clusters at 
cosmologically significant redshift. 
By comparing the bolometric luminosity of \ms with the best fit relationship 
log(L$_{BOL,X})=(2.88\pm0.15)$log(T/6keV)$+$(45.06$\pm$0.03) by Arnaud \& 
Evrard (1999), or using similar observationally determined relationships,
one would expect for \ms a temperature T$=$8.2$\pm$0.15 keV which is
well in agreement within the errors with the measured XMM-{\it Newton} 
temperature. 

We check next the $\sigma_{V}-T_{X}$ relationship. 
We can estimate the velocity 
dispersion for  \ms from the X-ray data and compare the result with the 
optical data. The velocity dispersion implied by the XMM-{\it Newton} 
temperature   using the relation T$_{X}/\mu m_{p} \sigma^{2}$  with 
$\beta=$1 (that is assuming energy equipartition between the 
galaxies and the gas) is   $\sigma_{V}=1047^{+50}_{-44}$ km s$^{-1}$,  
consistent with the observed  optical velocity  dispersion within the errors,
implying that the velocity dispersion reflects the temperature of the gas. 
Girardi et al. (1996) have derived a best fit relation between the 
velocity dispersion and the X-ray temperature equal to
Log($\sigma$)$=$(2.53$\pm$0.04)+(0.61$\pm$0.05)log(T),
with more than 30\% reduced scatter with respect to previous  work by 
taking into account distortions in the velocity fields, asphericity of 
the cluster or presence of substructures. Using the temperature T$=$7.2 
keV of \ms in the above relation, the resulting velocity dispersion value, 
$\sigma_{V}=1129^{+66}_{-57}$ km s$^{-1}$,  is in very good agreement with 
the measured optical value,  $\sigma_V=1153\pm$80 km s$^{-1}$. 
The relation between temperature and velocity dispersion
has been published by several authors. Last in order of time, this 
relation can be found in Lubin, Mulchaey \& Postman (2003) where the 
authors have collected the information known today for all clusters 
at redshfits of z$\leq$1.27 with a measured X-ray temperature.
There is a big scatter in the $\sigma_{V}-$T$_{X}$ relation (see their 
figure 4) but the average relation is consistent with 
$\sigma \propto T^{1/2}$ up to these  redshifts. We find that \ms ~follows 
closely this correlation.

We can also estimate the mass of \ms using the X-ray temperature value.
With the assumptions that the mean density in the virialized region is 
$\sim$200 times the critical density at the redshift of the cluster and that 
the cluster is  isothermal (Evrard, Metzler \& Navarro 1996; Donahue et al. 
1998), we can use the  scaling law method as illustrated in  Arnaud \& Evrard 
(1999).  From the simulations of Evrard, Metzler \& Navarro (1996) for the 
mass-temperature relation one can estimate the virial mass within a radius
r$_{200}(T)=$1.85(T/10keV)$^{1/2}(1+z)^{-3/2}$ h$^{-1}$ Mpc by using the  
equation  M$_{vir}\approx(1.45\times10^{15} h^{-1} M_{\sun})(1+z)^{-3/2}(kT_{X}/10 keV)^{1.5}$.  
From the XMM-{\it Newton} temperature the virial mass is approximately
M$_{vir}=(3.58^{+0.53}_{-0.44})\times10^{14}$ M$_{\sun}$ ~within 
r$_{200}=0.63 ~h^{-1}$ Mpc, where the uncertainties on the mass reflect the 
uncertainties on the temperature. This mass is  lower than the one derived by 
J01 $(M\leq 0.76 ~h^{-1}_{100}$ Mpc = 6.2$\times10^{14} ~h^{-1}_{100} ~M_{\sun}$)
because they use a higher Chandra temperature which, given the relation 
between r$_{200}$ and T, converts in a larger r$_{200}$ (their 0.76 
$h^{-1}_{100}$ Mpc vs our 0.63 $h^{-1}_{100}$ Mpc). 
The virial mass derived from the XMM-{\it Newton} temperature is  definitely
lower than the mass derived from weak lensing by Hoekstra, Franx \&  
Kuijken (2000) of 1.2$\pm$0.2$ \times10^{15}$ $h^{-1}_{50} M_{\sun}$ who 
used a different aperture radius of 1~$h^{-1}_{50}$ Mpc. It is not possible
with the present XMM-{\it Newton} data to assume an aperture radius of  
1~$h^{-1}_{50}$ Mpc (1.75\arcmin ~using the Hoekstra et al. cosmology) 
since we are running into the gaps between the chips for the pn detector.

 \begin{figure}
   \centering
     \caption{21,600s I-band image of MS\,1054$-$0321 taken at the 
            2.2m of the University of Hawai'i. The smoothed overlayed
            X-ray contours are in units of 3, 5, 7, 10, 15, 20, 30, 35, 40, 
            43, 45 $\sigma$ over the background. The image measures 
            $2048 \times 2048$  pixels covering a field of 
            $7'.5\times7'.5$ ($1.86h\times1.86h$ Mpc at $z =$ 0.83).
            Figure 7 can be seen in the pdf file of the paper at
             http://www.ira.cnr.it/~gioia/PUB/publications.html}
         \label{Fig7}
   \end{figure}
Given the XMM-{\it Newton} results, \ms appears to be a more ``normal'' 
cluster with high, but not extreme  X-ray luminosity or temperature values. 
Its  temperature seems to be commensurate with the predictions from
its X-ray luminosity from the L$_{X}-T_{X}$ relation of local clusters 
published in the literature.  The optically measured velocity dispersion 
is consistent with the velocity dispersion expected from the 
$\sigma_{V}-T_{X}$  relationship.  Given the clear 
evidence for substructure in the cluster morphology, indicative of a 
still unrelaxed cluster, it is surprising that the global  properties
of \ms fit so well in any  local cluster correlation. 

\section{Summary}

We have presented new observations for the z$=$0.83 galaxy cluster \ms 
performed with the instruments on board the XMM-{\it Newton} 
satellite. The main result of this paper is the lower value indicated
by the XMM-{\it Newton} observations for the overall cluster X-ray  
temperature with respect to some of the previous analyses. 
Excluding the contribution  of the southern  point source the temperature 
of the cluster is T$=$7.2$^{+0.7}_{-0.6}$ keV.  If one allows the column
density to be a free parameter then the best fit temperature is
T$=$8.6$^{+1.2}_{-1.1}$ keV,  and the best fit column density becomes 
N$_{H}=1.33^{+0.15}_{-0.14}\times10^{20}$ atoms cm$^{-2}$. 
The iron line  is well  detected with a value for  the abundance of 
Z$=$0.33$^{+0.19}_{-0.18}$ Z$_{\sun}$. The point source to the south is  
fitted by an absorbed power law with  N$_{H}=(1.7\pm0.7)\times10^{21}$ 
atoms cm$^{-2}$ and a photon index $\Gamma$=1.9$\pm$0.2. 
The XMM-{\it Newton} data confirm the  substructure in the cluster  
X-ray morphology already  seen  by  previous  experiments.  The two 
main clumps have very  different temperatures. The eastern clump is 
hotter (T$=$8.1$^{+1.3}_{-1.2}$ keV) compared to the 
western clump  (T$=$5.6$^{+0.8}_{-0.6}$ keV), as also found by J01. 
\ms has an X-ray luminosity  in the  2--10 keV energy band of
$L_{X}=(3.81\pm0.19) \times10^{44}$ h$^{-2}$ erg s$^{-1}$ and
a  bolometric luminosity of
$L_{BOL}=(8.05\pm0.40) \times10^{44}$ h$^{-2}$ erg s$^{-1}$.
The velocity  dispersion derived from the X-ray temperature  for the whole 
cluster is in good agreement with the most recent  determination of the 
observed  velocity  dispersion derived from a much enlarged database of 145 
cluster member redshifts. \ms fits well in the  $\sigma \propto T^{1/2}$
correlation available in the literature and derived from information
collected for all clusters  at redshfits of z$\leq$1.27 known today and
with a measured X-ray temperature.

We have discussed in detail the discrepancy between the XMM-{\it Newton}  
and Chandra derived temperatures and found that the origin of this 
discrepancy is mainly due to corrections to be implemented in the
Chandra data to obtain a much cleaner dataset. 
An  analysis of the Chandra archival data taking into account all the 
possible corrections yields a temperature T$=$7.4$^{+1.4}_{-0.9}$ keV,
in excellent agreement with the XMM-{\it Newton} determined 
temperature value. Given these results \ms  is no longer amongst the 
hottest clusters known but it is  more similar  to other X-ray selected 
clusters in the same redshift range  with moderate temperatures and 
elongated or double morphologies seen in different waveband domains.

\begin{acknowledgements}
We are grateful to G. Illingworth and M. Franx for access to the \ms
Keck spectroscopic database, and to the referee, M. Donahue, who made
many constructive criticisms which improved this paper. 

\end{acknowledgements}

\newpage
%
%
%
%
%
%
%
%


\begin{table*}
\begin{center}

\caption{X-RAY PROPERTIES OF \ms}

\begin{tabular}{llllrcccl}

& & & & & & & & \\

\hline
\hline
           &             &       &   &            &    &    &  & \\
RA (J2000) & DEC (J2000) & Instr & R & Net Cts & T  & Z  & N$_{H}$ & Region \\
 ~~h ~~m ~~s & ~~~\deg ~~~\arcmin ~~~\arcsec &  & \arcmin & 2$-$10 keV & keV & Z$_{\sun}$ & $\times10^{20}$ at/cm$^{2}$& \\

\hline

& & & & & & & & \\

10~56~58.7 & -03~37~45 & &  & & 7.2$^{+0.7}_{-0.6}$ & 0.33$^{+0.19}_{-0.18}$ & frozen$^{\mathrm{a}}$ &  whole cluster \\
& & &  & & 8.6$^{+1.2}_{-1.1}$ & 0.22$^{+0.15}_{-0.14}$ & 1.33$^{+0.15}_{-0.14}$ &  whole cluster \\

   & & MOS1 & 1.5 & 2198$\pm$52 & &  &  & \\
   & & MOS2 & 1.5 & 2392$\pm$54 & &  &  & \\
   & & pn   & 1.5 & 4486$\pm$75 & &  &  & \\
10~57~00.3 & -03~37~29 & &  & & 8.1$^{+1.3}_{-1.2}$& 0.12$^{+0.35}_{-0.12}$ & frozen$^{\mathrm{a}}$ & Component~1 \\
   & & MOS1 & 0.47 & 583$\pm$25  & &  &  & \\
   & & MOS2 & 0.47 & 604$\pm$25  & &  &  & \\
   & & pn   & 0.47 & 1200$\pm$36 & &  &  & \\
10~56~56.6 & -03~37~36 & & & & 5.6$^{+0.8}_{-0.6}$ & 0.51$^{+0.36}_{-0.32}$ & frozen$^{\mathrm{a}}$ & Component~2 \\
   & & MOS1 & 0.47 & 524$\pm$24  & &  & & \\
   & & MOS2 & 0.47 & 575$\pm$25 & &  &  &\\
   & & pn   & 0.47 & 1146$\pm$36 & &  & & \\

& & & & & & & \\

\hline
\hline

\end{tabular}
\end{center}
\begin{list}{}{}
\item[$^{\mathrm{a}}$]  N$_{H}$ fixed at the Galactic value of $3.6\times10^{20}$ 
atoms cm$^{-2}$
\end{list}

\end{table*}
\label{tab1}
%


\begin{table*}
\begin{center}

\caption{FLUXES AND LUMINOSITIES FOR THE CLUSTER COMPONENTS}

\begin{tabular}{llcccl}

& & & & & \\

\hline
\hline
           &             &       &   &  &   \\
RA (J2000) & DEC (J2000) & F$_{X}$ (2--10 keV) & L$_{X}$ (2--10 keV) & L$_{BOL}$ & Region \\
 ~~h ~~m ~~s & ~~~\deg ~~~\arcmin ~~~\arcsec & $\times10^{-13}$ erg cm$^{-2}$ s$^{-1}$ & $\times10^{44}$ erg s$^{-1}$& $\times10^{44}$ erg s$^{-1}$  & \\

\hline

& & & & & \\
10~56~58.7 & -03~37~45 & 3.40$\pm$0.17 & 3.81$\pm$0.19 & 8.05$\pm$0.40 & whole cluster \\
10~57~00.3 & -03~37~29 & 0.97$\pm$0.07 & 1.05$\pm$0.07 & 2.23$\pm$0.15 & eastern Component~1 \\
10~56~56.6 & -03~37~36 & 0.69$\pm$0.06 & 0.87$\pm$0.07 & 1.82$\pm$0.15 & western Component~2 \\

& & & & & \\

\hline
\hline

\end{tabular}
\end{center}

\end{table*}
\label{tab2}


\begin{table*}
\begin{center}

\caption{X-RAY PROPERTIES OF THE POINT SOURCE TO THE SOUTH}

\begin{tabular}{llllrccc}

& & & & &  & & \\

\hline
\hline
           &             &       &   &            &    & &   \\
RA (J2000) & DEC (J2000) & Instr & R & Net Cts & N$_{H}$  & $\Gamma$ & Flux (2--10 keV) \\
 ~~h ~~m ~~s & ~~~\deg ~~~\arcmin ~~~\arcsec & & \arcmin & & $\times10^{21}$ at/cm$^{2}$ & & $\times10^{-14}$ erg cm$^{-2}$ s$^{-1}$\\

\hline

& & & & & & & \\
10~56~58.8 & -03~38~52 & &  & & 1.7$\pm$0.7 & 1.9$\pm$0.2 & 3.0$\pm$1 \\
   & & MOS1 & 0.33 &  101$\pm$11 & &    & \\
   & & MOS2 & 0.33 &   89$\pm$11 & &    & \\
   & & pn   & 0.33 &  264$\pm$18 & &    & \\
\\
\hline
\hline

\end{tabular}
\end{center}
\end{table*}
\label{tab3}


\begin{table*}
\begin{center}

\caption{OPTICAL AND X-RAY PROPERTIES OF THE TWO CLUSTER COMPONENTS}

\begin{tabular}{llllcccc}

&&&&&&&\\

\hline
\hline
           &             &   & &      &             &  &    \\
RA (J2000) & DEC (J2000) & R & N$_{gals}$ & $<z>$ & $\sigma_V$  & T$_{X}$ & L$_{BOL}$\\
 ~~h ~~m ~~s  & ~~~\deg ~~~\arcmin ~~~\arcsec & ~\arcmin & & & km s$^{-1}$ & keV & 10$^{44}$ erg s$^{-1}$ \\

\hline
& & & & & & &\\
10~57~00.3 & -03~37~28.8 & 0.47 & 17& 0.8297$\pm$0.0020 & 1225$\pm$250 & 8.1 & 2.23 \\
10~56~56.6 & -03~37~36.3 & 0.47 & 11&0.8342$\pm$0.0022 & 1353$\pm$315  & 5.6 & 1.82 \\
&&&&&&&\\

\hline
\hline

\end{tabular}
\end{center}
\end{table*}
\label{tab4}

\end{document}